\font\bfgr=eurb10 

\def\frac#1#2{{#1\over#2}} 
\def\fract#1#2{{\textstyle {#1\over#2}}}
\def\newline{\hfill\break}

\def\ip#1#2{\langle #1,#2\rangle} 
\def\expc#1{\langle #1\rangle} 
\def\conj#1{\overline{#1}}
\def\tr{{\rm tr}}
\def\car{{\rm CAR}}

\def\real{{\bf R}}

\def\balpha{\hbox{{\bfgr\char11}}}
     
\def\pd#1#2{\frac{\partial #1}{\partial #2}}    
\def\sec#1{\advance\count31 by 1\bigskip\noindent{\bf \the\count31. #1}
\medskip}
\def\im#1{\widetilde{#1}}
\def\ref{\advance\count32 by 1\item{[\the\count32]}}

\def\im#1{\widetilde{#1}}

\count31=0
\count32=0

\centerline{\bf Fermion mixing in quasifree states}

\bigskip
\centerline{K.C.\ Hannabuss and D.C.\ Latimer}
\medskip
\centerline{Mathematical Institute, 24-29, St Giles', Oxford OX1 3LB}

\bigskip\noindent{\bf Abstract}
Quantum field theoretic treatments of fermion oscillations are typically 
restricted to calculations in Fock space.
In this letter we extend the oscillation formulae to include more general 
quasi-free states, and also consider the case when the mixing is not
unitary.

\sec{Introduction}

The theoretical underpinning for fermion oscillations was developed
decades ago [6], but only recently have terrestrial and solar neutrino
experiments begun to substantiate this work [22,23], and experimental
evidence for oscillations has led to renewed interest in the theory.
Current quantum field theoretic treatments of the fermion oscillation 
phenomena [1,2,7-18,21] have brought about modifications to the
oscillation formula developed in [6], adding terms depending on the sum
of energies rather than their difference.
Although these new results are based on calculations made in fermion Fock 
space, they rely on modifications suggested by physical considerations.
In this paper we generalise the oscillation formula to general quasifree 
states, and show that the additional terms occur naturally in that
setting.
Our result contains the known formula for Fock states as a special case,
but also includes other physical scenarios such as the thermal
(KMS) state, or situations in which polarisation of the vaccum has occurred.
The oscillation formula for the thermal state could be a better
approximation for fermions at  nonzero temperature such as solar
neutrinos.
At the end of the paper we shall also consider the case when the mixing 
operator is not unitary and show that it leads to similar effects.

From a mathematical point of view the main obstacle to such calculations
lies in the fact that we wish to calculate the oscillatory behaviour of
correlations between projections onto flavour states (such as the $\nu_e$
and $\nu_\mu$ states) at different times, but the dynamical behaviour is
simplest in states with definite masses. These are distinct from the
states of definite flavour to which they are related by a non-trivial {\it
mixing} transformation.

The fermionic anticommutation relations (CAR) can be written in terms of 
smeared creation operators 
$$c(w) = \int w(x)a^*(x)\,dx$$
(smeared with  test functions $w,z$ in the complex inner product space of
wave  functions on $\real^3$ with values in the product of Dirac spinors
${\cal V}$ and an $N$-dimensional space $V$ describing the various flavour
states) as
$$[c(w)^*,c(z)]_+ = \ip{w}{z}, \qquad [c(w),c(z)]_+ = 0.$$
It is well-known [19] that the (abstract) algebra defined by these
relations  has many inequivalent representations by operators in Hilbert
space, and the  study of the interrelations between a selection of these
forms the main focus of this paper.
We shall mainly be concerned with quasi-free representations of the CAR 
algebra which generalise the standard Fock and Dirac-Fock representations.
(The Fock representations themselves have been studied, for example, in 
[7,8,20,24].) 
Quasi-free representations are those in which the Wick determinant formula 
expresses the $n$-point correlation functions in terms of the 2-point 
correlation functions just as in Fock space.
As well as appearing for thermal states of systems, quasi-free states
often arise in situations where the vacuum is polarised, and so allow us
to treat more complicated field-theoretic effects whilst avoiding the
detailed models.
A well-known technique of Powers and St\o rmer [25] and Araki [3] tells us 
how to construct any quasi-free representation of $\car(W)$ as the
composition of an injection of $W$ into $W\oplus W$ with a Fock
representation of $\car(W\oplus W)$, and so we shall concentrate on Fock
representations.

\sec{The one particle space}

To establish notation we first recall that for Dirac particles the
elements of $W=W_m$ can be thought of as the initial data for the Dirac
equation 
$$i\hbar \pd{w}{t} = H_D w$$
where the Dirac Hamiltonian $H_D$ is given in terms of momentum operators
${\bf P}$ by
$$H_D = c(\balpha\otimes {\bf P} + \beta\otimes Mc)$$
with $\balpha = (\alpha_1,\alpha_2,\alpha_3)$ and  $\beta$ satisfying the 
Clifford algebra relations
$$\alpha_j\beta + \beta \alpha_j = 0,  \qquad \beta^2 = 1, \qquad
\alpha_j\alpha_k + \alpha_k\alpha_j = 2\delta_{jk}, \qquad j,k = 1,2,3,$$
and $M$ a positive operator on $V$ with eigenvalues the masses $m_1,m_2, 
\ldots, m_n$.
(In what follows, we shall omit the tensor produt and write $\beta M$ for 
$\beta\otimes M$, etc.)

Choosing a basis in which the mass matrix $M$ is diagonal, we refer to the 
solutions of the Dirac eqaution as mass eigenstates.
We define
$$P_\pm = \fract12\left(1\pm H_DE^{-1}\right),$$
where $E$ is the positive square root of the positive operator 
$H_D^2 = (|{\bf P}|^2 + M^2c^4)$. (We shall also use $E_j$ for the value
when 
in the eigenstate with mass $m_j$.)
The $P_\pm$ are idempotent, self-adjoint, and $P_\pm H_D = EP_\pm$; that
is, they are the positive and negative energy projections on $W$.
They also determine the {\it mass representation} of the CAR algebra with 
creation operators $c_m$ and  a Dirac-Fock vacuum vector $\Omega_m$ which 
satisfies the Dirac condition that, for every $w$ in $W$,
$$c_m(P_+w)^*\Omega_m = 0 = c_m(P_-w)\Omega_m.$$

When we study the flavour space $W_f$ we work rather with a {\it flavour 
representation} $c_f$ in a standard Fock space with a flavour vacuum
$\Omega_f$ satisfying $c_f(w)^*\Omega_f = 0$ for all $w\in W_f$.

\sec{General mixing transformations}

In our previous paper we studied what happened when the mass and flavour 
spaces $W_m$ and $W_f$ were isomorphic by a unitary mixing transformation
$T$, but here we shall consider more general situations such as orthogonal
mixing transformations, and Powers-St\o rmer transformations which enable
us to realise quasi-free states on Fock spaces.
When $T$ is unitary the spaces $W_m$ and $W_f$ can be identified, but, for 
orthogonal $T$, when $W_m$ and $W_f$ are the same as real spaces but have 
different complex structure, it is simpler to treat them as distinct.

It will be convenient to consider more generally the case when we have two 
inner product spaces $W_j$ and $W_k$ and a map $T_{jk}: W_k \to W_j$
which is orthogonal in the sense that it preserves the real part of the
inner product: for all $w$, $z \in W_k$, 
$$\ip{z}{w} + \ip{w}{z} = \ip{T_{jk}z}{T_{jk}w} + \ip{T_{jk}w}{T_{jk}z}.$$
For any real-linear operator $T_{jk}$ on $W_k$ we define the complex
linear map $a_{jk} = \frac12(T_{jk}-J_jT_{jk}J_k)$ and the antilinear map
$b_{jk} = \frac12(T_{jk}+J_jT_{jk}J_k)$ where $J_j$ and $J_k$ simply
indicate multiplication by $i$ on $W_j$ and $W_k$, respectively.
The Fock space creation and annihilation operators $c_j$ and $c_k$ are 
linked by
$$c_k(w) = c_j(a_{jk}w) + c_j(b_{jk}w)^*,$$
and
$$c_j(T_{jk}w) + c_j(T_{jk}w)^* 
= c_j(a_{jk}w) + c_j(b_{jk}w)^* + c_j(a_{jk}w)^* + c_j(b_{jk}w),$$
for all $w\in W_k$, where $a_{jk}$ and $b_{jk}$ are the Bogoliubov maps
just defined.
To be consistent with the anticommutation relations in both $W_j$ and
$W_k$ we have the orthogonality relations 
$$a_{jk}^*a_{jk} + b_{jk}^*b_{jk} = 1 = a_{jk}a_{jk}^* +
b_{jk}b_{jk}^*,\qquad
a_{jk}^*b_{jk} + b_{jk}^*a_{jk} = 0 = a_{jk}b_{jk}^* + b_{jk}a_{jk}^*.$$
These are just the conditions that $T$ be orthogonal.
Since these transformations mix creation and annihilation operators it is 
expedient to introduce a more succinct notation.
We combine creators and annihilators in the row vector $\im{c}_j = 
\left(\matrix{c_j &c_j^*}\right)$, and introduce 
$$\Lambda_{jk} = \left(\matrix{a_{jk} &b_{jk}\cr b_{jk}
&a_{jk}\cr}\right),$$
to obtain
$$\im{c}_k(\im{w}) = \im{c}_j(\Lambda_{jk}\im{w}), \qquad{\rm for}\qquad
\im{w} = \left(\matrix{w_1\cr w_2\cr}\right).$$

One advantage of dealing with the general situation is that there is an 
obvious composition law
$$\Lambda_{jl} = \Lambda_{jk}\Lambda_{kl},$$
obtained by writing the relationship between creation operators on $W_j$
and $W_l$ directly and through the intermediate space $W_k$.
The orthogonality properties of the Bogoliubov maps can also be
interpreted as telling us that $a_{jk} = a_{kj}^*$, and $b_{jk} =
b_{kj}^*$, or 
$$\Lambda_{jk} = \Lambda_{kj}^*.$$
(This is closely related to Araki's self-dual construction [3].)

Instead of the vacuum states described above we shall use a more general
quasi-free state $\omega$ for the CAR algebra.
As noted above, this is determined by the two-point correlation functions 
which define a complex linear operator $R$ and a conjugate linear operator 
$S$ by 
$$\eqalign{\omega[c(w)^*c(z)] &= \ip{w}{Rz}\cr
\omega[c(w)c(z)] &= \ip{Sw}{z}\cr}$$
where $R= R^*$, $0\leq R\leq 1$, and $S = -S^*$, [3,4,5].
We note that if $K = i(2R-2S-1)$ defines a complex structure on $W$, that
is $K^2 = -1$, then the state $\omega$ is a Fock state for some choice of 
complex structure.

The GNS construction guarantees the existence of a representation $\pi: 
\car(W) \to {\cal H}$ containing a cyclic vector $\Omega\in 
{\cal H}$ such that $\omega(b) = \ip{\Omega}{\pi(b)\Omega}$ for any 
$b\in \car(W)$.
The representation $\pi$ over ${\cal H}$ can be expressed in terms of 
a Dirac--Fock representation of $\car(W_+\oplus W_-)$, where $W_+$ and
$W_-$ are both isomorphic to $W$. 
The representation in question takes $(w_+,w_-)\in W_+\oplus W_-$ to 
$c_+(w_+) +c_-(w_-)$ and the $w_+$ with a Dirac type vacuum $\Omega$ which
is killed by the annihilators $c_+(w_+)^*$, and the creators $c_-(w_-)$.
This means that
$$\ip{c_+(w)\Omega}{c_+(z)\Omega} = \ip{w}{z}, \qquad
\ip{c_-(w)^*\Omega}{c_-(z)^*\Omega} = \ip{z}{w},$$
and all other two-point correlation functions vanish.
Suppressing the representation map, and writing $c = \pi\circ c$,  the
required representation of $\car(W)$ is  given by
$$c(w) = \fract{1}{\sqrt{2}}\left(c_+(a_{+}w)+c_+(b_{+}w)^* 
+ c_-(a_{-}w) + c_-(b_{-}w)^*\right)$$
where $a_{\pm}$ are linear and $b_{\pm}$ are antilinear and 
satisfy the orthogonality relations $a_{\pm}^*a_{\pm} + b_{\pm}^*b_{\pm} =
1$, $a_{\pm}^*b_{\pm} + b_{\pm}^*a_{\pm} = 0$.
(It is easy to check that this does provide a representation of
$\car(W)$.)

When $W= W_j$ we shall write $a_{\pm j}$ and $b_{\pm j}$ for the
Bogoliubov maps to give
$$c_j(w) = \fract{1}{\sqrt{2}}\left(c_+(a_{+j}w)+c_+(b_{+j}w)^* 
+ c_-(a_{-j}w) + c_-(b_{-j}w)^*\right).$$
We adopt the summation convention that when an index $s$ can range over 
different values $\{+,-\}$ and is repeated (as in expressions such as 
$c_s(a_{sj}w)$ or $c_j(a_{js}a_{sk}w)$) one sums over all its values and 
divides by $\sqrt{2}$.
Then the above expression can be abbreviated to 
$\im{c}_j(\im{w}) = \im{c}_s(\Lambda_{sj}\im{w})$, 
and it is easy to see that our earlier rules for compositions apply.
Thus a state which gives a quasi-free representation for the label $j$
will also do so for the label $k$.

With this notation we have $\im{c}_j(w) = \im{c}_s(\Lambda_{sj}w)$.
When applied to $\Omega$, only certain of the components are non-zero.
For instance,
$$c_j(w)\Omega = \frac1{\sqrt{2}}\left(c_+(a_{+j}w) + c_+(b_{+j}w)^*  
+ c_-(a_{-j}w) + c_-(b_{-j}w)^*\right)\Omega
= \frac1{\sqrt{2}}\left(c_+(a_{+j}w) + c_-(b_{-j}w)^*\right)\Omega,$$
and, similarly, 
$$c_j(w)^*\Omega = \frac1{\sqrt{2}}\left(c_+(b_{+j}w)  
+ c_-(a_{-j}w)^*\right)\Omega.$$
Reversing the expansion we obtain
$$c_j(w)\Omega 
= \frac1{2}\left(c_j((a_{j+}a_{+j}+b_{j-}b_{-j})w) 
+ c_j((a_{j-}b_{-j}+b_{j+}a_{+j})w)^*\right)\Omega,$$
and 
$$c_j(w)^*\Omega 
= \frac1{2}\left(c_j((a_{j+}b_{+j}+b_{j-}a_{-j})w)  
+ c_j((a_{j-}a_{-j}+b_{j+}b_{+j})w)^*\right)\Omega,$$
which may be combined as
$$\im{c}_j(\im{w})\Omega = \im{c}_j(\Lambda^0_{j}\im{w})\Omega,
\qquad{\rm where}\qquad
\Lambda^0_{j} = \left(\matrix{R &S^*\cr S &R^\prime\cr}\right),$$
where $R = \frac12(a_{j+}a_{+j}+b_{j-}b_{-j})$, 
$S = \frac12(a_{j+}b_{+j}+b_{j-}a_{-j})$, 
$S^* = \frac12(a_{j-}b_{-j}+b_{j+}a_{+j})$
(which the orthogonality relations show is the adjoint of $S$), and 
$R^\prime = \frac12(a_{j-}a_{-j}+b_{j+}b_{+j}) = 1-R$ (by the 
orthogonality relations).
In effect $\Lambda^0_{j}$ is the projection on $W_j$ associated with 
$\Omega$.
We easily calculate that
$$\ip{c_j(z)\Omega}{c_j(w)\Omega} 
= \ip{c_j(z)\Omega}{(c_j(Rw) + c_j(Sw)^*)\Omega}
= \ip{z}{Rw}$$
so that $R$ is the correlation operator already introduced, and similarly
for $S$.

Whenever $U$ is a unitary transformation of $W_j$, there is a unitary map 
$\pi_j(U)$ implementing $U$ in the sense that
$$\pi_j(U)c_j(w)\pi_j(U)^{-1} = c_j(Uw),$$
for all $w\in W_j$.
Since $\pi_j(U)$ is unitary there is a similar relation for the
annihilator 
$c_j(w)^*$ and we combine these as 
$$\pi_j(U)\im{c}_j(\im{w})\pi_j(U)^{-1} = \im{c}_j(\im{U}\im{w}),$$
where
$$\im{U} = \left(\matrix{U &0\cr 0 &U\cr}\right).$$
(When $j$ refers to flavour this gives a representation of the group of 
flavour transformations.)

We shall also need the infinitesimal version of this which arises from
taking the unitary $\exp(isP)$ and differentiating at $s=0$ to obtain
$$[\pi^\prime_j(P),c_j(w)] =
-i\frac{d}{ds}\left.\pi_j(e^{isP})c_j(w)\pi_j(e^{isP})^{-1}\right|_{s=0} 
= -i\frac{d}{ds}\left.c_j(e^{isP}w)\right|_{s=0}= c_j(Pw).$$
When $P = 1$, we can regard $N= \pi_f^\prime(1)$ as the number operator,
and when 
$P=P^\lambda$, the projection onto the states of flavour $\lambda$, then 
$N_\lambda = \pi_f^\prime(P^\lambda)$ counts the number of flavour
$\lambda$  particles.

For our calculations we need to know the effect of $\pi_j(U)$ on $c_k(w)$
for different $j$ and $k$.
We therefore note that 
$$\eqalign{\pi_j(U)\im{c}_k(\im{w})\pi_j(U)^{-1} &=
\pi_j(U)\im{c}_j(\Lambda_{jk}\im{w})\pi_j(U)^{-1}\cr
&= \im{c}_j(\im{U}\Lambda_{jk}\im{w})\cr
&= \im{c}_k(\Lambda_{kj}\im{U}\Lambda_{jk}\im{w})\cr
&= c_k\left((a_{kj}Ua_{jk}+b_{kj}Ub_{jk})w\right) 
+ c_k\left((b_{kj}Ua_{jk}+a_{kj}Ub_{jk})w\right)^*.\cr}$$ 
It is convenient to introduce the abbreviation 
$\im{U}_{k} = \Lambda_{kj}\im{U}\Lambda_{jk}$, and write
$$\pi_j(U)\im{c}_k(\im{w})\pi_j(U)^{-1} = \im{c}_k(\im{U}_{k}\im{w}),$$
(it being understood that $U$ is an operator on $W_k$).
We have the explicit formula
$$\im{U}_{k} = 
\left(\matrix{u_{k} &v_{k}\cr v_{k} &u_{k}}\right),$$
with 
$$u_{k} = a_{kj}Ua_{jk}+b_{kj}Ub_{jk}, \qquad
v_{k} = a_{kj}Ub_{jk}+b_{kj}Ua_{jk}.$$

\sec{Correlation functions}

The correlation functions which we first wish to calculate have the form
$$\ip{c_f(z)\Omega}{\pi_m(U)^*\pi_f^\prime(P)\pi_m(U)c_f(w)\Omega},$$
where $P$ projects onto a flavour state, $U = \exp(-itH_D/\hbar)$ gives 
the time evolution, and $\Omega$ defines a quasi-free state.
This can easily be found by differentiating the more tractable
$$\ip{c_f(z)\Omega}{\pi_m(U)^*\pi_f(D)\pi_m(U)c_f(w)\Omega},$$
where $D = \exp(isP)$.

Gathering together our various comments we calculate that
$$\eqalign{\pi_m(U)^*\pi_f(D)\pi_m(U)\im{c}_f(\im{w})\Omega
&= \pi_m(U)^*\pi_f(D)\pi_m(U)\im{c}_f(\Lambda^0_{f}\im{w})\Omega\cr
&= \im{c}_f(\im{U}^*_{f}\im{D}\im{U}_{f}\Lambda^0_{f}\im{w})
\pi_m(U)^*\pi_f(D)\pi_m(U)\Omega.\cr}$$
We now differentiate this to get
$$\eqalign{i\pi_m(U)^*\pi_f^\prime(P)\pi_m(U)c_f(w)\Omega
&= \im{c}_f(\im{U}^*_{f}i\im{P}\im{U}_{f}\Lambda^0_{f}\im{w})\Omega
+ i\im{c}_f(\im{U}^*_{f}\im{U}_{f}\Lambda^0_{f}\im{w})
\pi_m(U)^*\pi_f^\prime(P)\pi_m(U)\Omega\cr
&=
\im{c}_f(\Lambda^0_{f}\im{U}^*_{f}i\im{P}\im{U}_{f}\Lambda^0_{f}\im{w})\Omega
+
i\im{c}_f(\Lambda^0_{f}\im{w})\pi_m(U)^*\pi_f^\prime(P)\pi_m(U)\Omega.\cr}$$

Inserting this expression into the inner product, but with 
$\im{w} = \left(\matrix{w\cr 0\cr}\right)$, we obtain 
$$\eqalign{
\ip{c_f(z)\Omega}{\pi_m(U)^*\pi_f^\prime(P)\pi_m(U)\im{c}_f(\im{w})\Omega}
&= -i\ip{c_f(z)\Omega}
{\im{c}_f(\Lambda^0_{f}\im{U}^*_{f}i\im{P}\im{U}_{f}\Lambda^0_{f}\im{w})
\Omega}\cr
&\qquad+\ip{c_f(z)\Omega}
{\im{c}_f(\Lambda^0_{f}\im{w})\pi_m(U)^*\pi_f^\prime(P)\pi_m(U)\Omega}.\cr}$$
In each inner product we take the creation operators from the right to an 
adjoint acting on the left.
There the factor of $\Lambda^0_{f}$ ensures that the adjoint annihilates 
$\Omega$, so that we simply get
$$\eqalign{
\ip{c_f(z)\Omega}{\pi_m(U)^*\pi_f^\prime(P)\pi_m(U)\im{c}_f(\im{w})\Omega}
&= -i\ip{[\im{c}_f(\Lambda^0_{f}\im{U}^*_{f}i\im{P}\im{U}_{f}\Lambda^0_{f}
\im{w})^*,c_f(z)]_+\Omega}{\Omega}\cr
&\qquad+\ip{[\im{c}_f(\Lambda^0_{f}\im{w})^*,c_f(z)]_+\Omega}
{\pi_m(U)^*\pi_f^\prime(P)\pi_m(U)\Omega}.\cr}$$
The anticommutators can now be written explicitly in terms of inner
products.
For example, the second gives
$$[\im{c}_f(\Lambda^0_{f}\im{w})^*,c_f(z)]_+
= [c_f(Rw)^*+c_f(Sw),c_f(z)]_+ = \ip{Rw}{z}.$$
The first requires a more detailed calculation, but we require only the
first entry in
$\Lambda^0_{f}\im{U}^*_{f}i\im{P}\im{U}_{f}\Lambda^0_{f}\im{w}$:
$$\eqalign{
\left(\matrix{1 &0\cr}\right)&\left(\matrix{R &S^*\cr S
&R^\prime\cr}\right)
\left(\matrix{u^*_{f} &v^*_{f}\cr 
v^*_{f} &u^*_{f}\cr}\right)
\left(\matrix{iP &0\cr 0 &iP\cr}\right)
\left(\matrix{u_{f} &v_{f}\cr 
v_{f} &u_{f}\cr}\right)
\left(\matrix{R &S^*\cr S &R^\prime\cr}\right)
\left(\matrix{w\cr 0\cr}\right)\cr
&= 
\left(\matrix{R &S^*\cr}\right)
\left(\matrix{u^*_{f} &v^*_{f}\cr 
v^*_{f} &u^*_{f}\cr}\right)
\left(\matrix{iP &0\cr 0 &iP\cr}\right)
\left(\matrix{u_{f} &v_{f}\cr 
v_{f} &u_{f}\cr}\right)
\left(\matrix{Rw\cr Sw\cr}\right).\cr}$$
Recalling that $v$ is conjugate linear, the product of the three middle 
matrices can be written as
$$
\left(\matrix{u^*_{f} &v^*_{f}\cr 
v^*_{f} &u^*_{f}\cr}\right)
\left(\matrix{iP &0\cr 0 &iP\cr}\right)
\left(\matrix{u_{f} &v_{f}\cr 
v_{f} &u_{f}\cr}\right)
= i\left(\matrix{F &G\cr G &F\cr}\right),$$
where 
$$F = u_f^*Pu_f-v_f^*Pv_f,\qquad
G = u_f^*Pv_f - v_f^*Pu_f.$$
Using the fact that $R=R^*$ and recalling the conjugate linearity of $S$,
this enables us to rewrite the first commutator as
$$\left(\matrix{R &S^*\cr}\right)
i\left(\matrix{F &G\cr G &F\cr}\right)
\left(\matrix{Rw\cr Sw\cr}\right)
= i\left(R^*FR +R^*GS - S^*GR -S^*FS\right)w.$$
Combining the expressions for the two commutators we obtain
$$\ip{z}{(R^*FR+R^*GS - S^*GR - S^*FS)w}
+\ip{z}{Rw}\ip{\Omega}{\pi_m(U)^*\pi_f^\prime(P)\pi_m(U)\Omega}.$$

The physically interesting quantity is the expected number of flavour 
$\lambda$ particles $\expc{N_\lambda(t)}_\mu$, in a state where one
flavour $\mu$ particle has been created out a time $t$ earlier out of the 
\lq\lq vacuum\rq\rq\ $\Omega$.
This can be obtained by taking  $P = P^\lambda$, $z = w = P^\mu\phi_j$, 
$U = V_t = \exp(-iH_D t/\hbar)$ above, and then summing as $\phi_j$ runs
over an orthonormal basis of $W$, to get (with $F^\lambda$ and
$G^\lambda$ denoting $F$ and $G$ when we take $P = P^\lambda$)
$$\eqalign{\sum_j
\ip{c_f(P^\mu\phi_j)\Omega}
{\pi_m(V_t)^*\pi_f^\prime(P)\pi_m(V_t)c_f(P^\mu\phi_j)\Omega}
&= \tr\left[\left(R^*F^\lambda R+R^*G^\lambda S - S^*G^\lambda R -
S^*F^\lambda 
S\right)P^\mu\right]\cr
&\qquad+\tr\left[RP^\mu\right] 
\ip{\Omega}{\pi_m(V_t)^*\pi_f^\prime(P^\lambda)\pi_m(V_t)\Omega}.\cr}$$
The last inner product is just the vacuum expectation of $P^\lambda$ and
should be subtracted (since we are only interested in the enhancement
produced by creating a flavour state), and we must also divide by the
norm of the state 
$\tr[RP^\mu]$ to get 
$$\expc{N_\lambda(t)}_\mu =
\frac{\tr\left[\left(R^*F^\lambda R + R^*G^\lambda S - S^*G^\lambda R 
- S^*F^\lambda S\right)P^\mu\right]}{\tr\left[RP^\mu\right]} .$$
The expected total flavour number $\expc{N(t)}_\mu$ is obtained by summing 
over $\lambda$, (which means that $F^\lambda$ and $G^\lambda$ are replaced
by 
$u_f^*u_f-v_f^*v_f$ and $u_f^*v_f - v_f^*u_f$, respectively.
The ratio $\expc{N_\lambda(t)}_\mu/\expc{N(t)}_\mu$ then gives the
proportion of flavour $\lambda$ particles.
In the next two sections we shall look at two special cases of this
formula.

During the preparation of this paper an interesting preprint appeared [17] 
which investigates the CP violation in three flavour mixing.
We note that in our context the transition probability for antiparticles
can  be calculated by using annihilation in place of creation operators,
which leads to replacement of $R$ by $R^\prime= 1-R$ and of $S$ by 
$S^* =-S$, to give
$$\expc{N_{\conj{\lambda}}(t)}_{\conj{\mu}} =
\frac{\tr\left[\left((1-R)^*F^\lambda (1-R)- (1-R)^*G^\lambda S 
+ S^*G^\lambda(1-R) - S^*F^\lambda S\right)P^\mu\right]}
{\tr\left[R^\prime P^\mu\right]}.$$

\sec{Unitary mixing in quasi-free states} 

The first case which we shall consider is for a unitary mixing matrix but
in a quasi-free state.
When the mixing transformation is given by a unitary operator $T$, we have 
$u_{f} = TUT^*$ and $v_{f} = 0$, so that $G^\lambda = 0$ and $F^\lambda = 
TV_t^*T^*P^\lambda TV_tT^*$.
The expression then simplifies to 
$$\expc{N_\lambda(t)}_\mu =
\frac{\tr\left[\left(R^*TV_t^*T^*P^\lambda TV_tT^*R 
- S^*TV_t^*T^*P^\lambda
TV_tT^*S\right)P^\mu\right]}{\tr\left[RP^\mu\right]} .$$
(This formula can easily be checked in the case of a thermal state 
$\omega_\beta$ at absolute temperature $(k\beta)^{-1}$, where $k$ is 
Boltzmann's constant (see the Appendix), and gives the known values 
$R = \left(1+e^{-\beta H_D}\right)^{-1}$ and $S=0$, in agreement with our 
general formula.)

We note that summation over $\lambda$ gives 
$$\expc{N(t)}_\mu = \frac{\tr\left[\left(R^*TV_t^*T^*TV_tT^*R 
- S^*TV_t^*T^*TV_tT^*S\right)P^\mu\right]}
{\tr[RP^\mu]} 
= \frac{\tr[(R^*R - S^*S)P^\mu]}{\tr[RP^\mu]},$$
which is, as one would hope,  independent of time

In fact for states invariant under global U(1) transformations of $W$, 
we always have $S=0$ [4,5], and then
$$\expc{N_\lambda(t)}_\mu =
\frac{\tr[R^*TV_t^*T^*P^\lambda TV_tT^*R P^\mu]}
{\tr[RP^\mu]}.$$

To achieve a more explicit formula we take $R = T\rho(H_D)T^*$ to be a 
function of the Hamiltonian, where $\rho(x)$ is a real function of the
real variable $x$, defined everywhere except, perhaps, zero.
Both the Dirac and KMS states satisfy this restriction.
This gives
$$\expc{N_\lambda(t)}_\mu= 
\frac{\tr[T\rho(H_D)V_t^*T^*P^\lambda TV_t\rho(H_D)T^*P^\mu]}
{\tr[T\rho T^*P^\mu]}
= \frac{\tr[\rho(H_D)V_t^*T^*P^\lambda TV_t\rho(H_D)T^*P^\mu T]}
{\tr[\rho T^*P^\mu T]}.$$
Performing some preliminary calculations, we have
$$\eqalign{\rho(H_D)V_t^* &= \rho(H_D)e^{iH_Dt/\hbar} (P_+ + P_-)\cr
&= \rho(E)e^{iEt/\hbar} P_+ + \rho(-E)e^{-iEt/\hbar}P_-\cr
&= \fract12\left[\rho(E)e^{iEt/\hbar} + \rho(-E)e^{-iEt/\hbar}\right]
+\fract12\left[\rho(E)e^{iEt/\hbar} 
- \rho(-E)e^{-iEt/\hbar}\right](P_+-P_-),\cr}$$
and, in particular,
$$T^*RT = \fract12\left[\rho(E) + \rho(-E)\right] 
+ \fract12\left[\rho(E) - \rho(-E)\right](P_+-P_-).$$

To condense our notation, we define
$$\eqalign{\sigma_j &= \rho(E_j)e^{iE_jt/\hbar} +
\rho(-E_j)e^{-iE_jt/\hbar}\cr
\delta_j &= \rho(E_j)e^{iE_jt/\hbar} - \rho(-E_j)e^{-iE_jt/\hbar}\cr
\gamma_j &= \rho(E_j) + \rho(-E_j)\cr
\epsilon_j &= c(\balpha\cdot{\bf P} + \beta m_jc)/E_j.\cr}$$
We note that $\epsilon_j$ has trace zero but
$$\tr(\epsilon_j\epsilon_k) = (|{\bf P}|^2c^2 + m_jm_kc^4)/(E_jE_k) =
S_{jk}.$$ 
We also know that in terms of the mass basis the components of $P^\lambda$
are
$$(T^*P^\lambda T)_{jk} = T_{j\lambda}^*T_{\lambda k} 
=  \conj{T}_{\lambda j}T_{\lambda k}.$$

With this notation the numerator in the oscillation formula is 
$$\eqalign{\fract14\sum_{j,k=1}^N&\tr[(\sigma_j+\delta_j\epsilon_j)T^*P^\lambda
T(\conj{\sigma}_k+\conj{\delta}_k\epsilon_k)^*T^*P^\mu T]
= \sum_{j,k=1}^N[\sigma_j\conj{\sigma}_k+\delta_j\conj{\delta}_kS_{jk}]
\conj{T}_{\lambda j}T_{\lambda k}\conj{T}_{\mu k}T_{\mu j}\cr
&=
\fract12\sum_{j,k=1}^N[(\sigma_j\conj{\sigma}_k+\delta_j\conj{\delta}_k)
(1+S_{jk})+(\sigma_j\conj{\sigma}_k-\delta_j\conj{\delta}_k)(1-S_{jk})]
\conj{T}_{\lambda j}T_{\lambda k}\conj{T}_{\mu k}T_{\mu j}.\cr}$$
Now, recalling that $\rho$ is a real function, we have
$$\eqalign{\sigma_j\conj{\sigma}_k+ \delta_j\conj{\delta}_k
&= 2[\rho(E_j)\rho(E_k)e^{i(E_j-E_k)t/\hbar}
+\rho(-E_j)\rho(-E_k)e^{i(E_k-E_j)t/\hbar}]\cr
\sigma_j\conj{\sigma}_k - \delta_j\conj{\delta}_k
&= 2[\rho(E_j)\rho(-E_k)e^{i(E_j+E_k)t/\hbar}
+\rho(-E_j)\rho(E_k)e^{-i(E_j+E_k)t/\hbar}].\cr}$$
When $m_j=m_k$ we have $S_{jk} = 1$ so that the second term in the
numerator disappears, and since also $E_j= E_k$,  we see that 
$\sigma_j\conj{\sigma}_k+\delta_j\conj{\delta}_k$ 
is time-independent, and consequently there is no flavour oscillation
between these.
For Fock states and general masses, one has $\rho(E)=1$ when $E\geq 0$,
and 
$\rho(E)=0$ when $E< 0$, which gives 
$\sigma_j\conj{\sigma}_k - \delta_j\conj{\delta}_k=0$, and 
$\sigma_j\conj{\sigma}_k+ \delta_j\conj{\delta}_k=
2e^{i(E_j-E_k)t/\hbar}$.

In general, the denominator is 
$$\tr\left[RP^\mu\right] = 2\sum_{j=1}^N \gamma_j\conj{T}_{\mu k}T_{\mu
j}.$$
Whenever $\rho(E)+\rho(-E)=1$, as happens for Fock states and also the
thermal states where $\rho(E) = (1+e^{-\beta E})^{-1}$, we have $\gamma_j =1$,
and then the unitarity of $T$ means that the denominator is 2, giving
$$\expc{N_\lambda(t)}_\mu =
\fract12{\sum_{j,k=1}^N[(\sigma_j\conj{\sigma}_k+\delta_j\conj{\delta}_k)
(1+S_{jk})+(\sigma_j\conj{\sigma}_k-\delta_j\conj{\delta}_k)(1-S_{jk})]
\conj{T}_{\lambda j}T_{\lambda k}\conj{T}_{\mu k}T_{\mu j}}$$
and our earlier formulae for 
$\sigma_j\conj{\sigma}_k \pm \delta_j\conj{\delta}_k$ show that this 
contains both standard oscillations depending on the energy differences
and others depending on $E_j+E_k$.
In the Fock vacuum state the oscillation formula is consistent with the 
calculations performed in [20].

Taking $\lambda =\mu$, the oscillation formula takes on a slightly more 
compact form
$$\eqalign{\expc{N_\mu(t)}_\mu &=
\fract12\sum_{j,k=1}^N\left[(\sigma_j\conj{\sigma}_k+\delta_j\conj{\delta}_k)
(1+S_{jk}) +
(\sigma_j\conj{\sigma}_k-\delta_j\conj{\delta}_k)(1-S_{jk})\right]
\left|T_{\mu j}\right|^2\left|T_{\mu k}\right|^2.\cr}$$

\sec{Non-unitary mixing in a Fock state}

We could instead work with $\Omega$ the flavour vacuum.
Then there is no need to inject $W$ into $W\oplus W$, so that we have
$S=0$ and $R=1$
This gives
$$\expc{N_\lambda(t)}_\mu =
\frac{\tr\left[F^\lambda P^\mu\right]}{\tr\left[P^\mu\right]},$$
where
$$F^\lambda = u_f^*P^\lambda u_f-v_f^*P^\lambda v_f,$$
with
$$u_{f} = a_{fm}V_ta_{mf}+b_{fm}V_tb_{mf}, \qquad
v_{f} = a_{fm}V_tb_{mf}+b_{fm}V_ta_{mf}.$$
From this we may show that both sorts of oscillation terms occur in this
case too.
However, the total flavour number is given by replacing $F^\lambda$ by 
$F = u_f^*u_f - v_f^*v_f$, and 
$$\expc{N_\lambda(t)}_\mu =
\frac{\tr\left[FP^\mu\right]}{\tr\left[P^\mu\right]},$$
and in general this depends on the time $t$.
This is essentially a squeezing phenomenon.
It provides a strong reason to be cautious about non-unitary mixing of
this kind.

\bigskip\noindent
{\bf Appendix}

For a thermal state at temperature $(k\beta)^{-1}$ the KMS condition and 
anticommutation relations give formally 
$$\eqalign{\omega_\beta[c(w)^*\im{D}c(z)]
&= \omega_\beta[\im{D}c(z)c(e^{\beta H_D}w)^*]\cr
&= \omega_\beta[\im{D}\left(\ip{e^{\beta H_D}w}{z}-c(e^{\beta 
H_D}w)^*c(z)\right)]\cr
&= \ip{e^{\beta H_D}w}{z}\omega_\beta[\im{D}]
- \omega_\beta[c(e^{\beta H_D}w)^*\im{D}c(z)]
+ \omega_\beta[c(De^{\beta H_D}w)^*c(z)].\cr}$$
This can be rearanged as
$$\omega_\beta[c((1+e^{\beta H_D})w)^*\im{D}c(z)]
= \ip{e^{\beta H_D}w}{z}\omega_\beta[\im{D}]
+ \omega_\beta[c(De^{\beta H_D}w)^*c(z)],$$
or, replacing $w$ by $(1+e^{\beta H_D})^{-1}w$, 
$$\omega_\beta[c(w)^*\im{D}c(z)]
= \ip{(1+e^{-\beta H_D})^{-1}w}{z}\omega_\beta[\im{D}]
+ \omega_\beta[c(D(1+e^{-\beta H_D})^{-1}w)^*c(z)].$$
The case $\im{D} =1$ (and $D=0$) gives the usual two point correlation 
function
$$\omega_\beta[c(w)^*c(z)]
= \ip{(1+e^{-\beta H_D})^{-1}w}{z},$$
so that
$$\omega_\beta[c(w)^*\im{D}c(z)]
= \ip{(1+e^{-\beta H_D})^{-1}w}{z}\omega_\beta[\im{D}]
+ \ip{(1+e^{-\beta H_D})^{-1}D(1+e^{-\beta H_D})^{-1}w}{z}.$$

\bigskip\noindent
{\bf Acknowledgements.}
We are grateful to M. Blasone and Tsou S.T. for useful conversations on 
different aspects of neutrino oscillation.
  
The second author would like to thank the Rhodes Trust for their support.

\bigskip\noindent
{\bf References}

\medskip
\ref E.\ Alfinito, M.\ Blasone, A.\ Iorio, and G.\ Vitiello,
\lq Squeezed Neutrino Oscillations in Quantum Field Theory\rq
{\it Phys. Lett. B},  hep-ph/9510213.
\ref E.\ Alfinito, M.\ Blasone, A.\ Iorio and G.\ Vitiello,
\lq Neutrino Mixing and Oscillations in Quantum Field Theory\rq, 
{\it Acta Phys.Polon. B} {\bf 27} (1996) 1493-1502, hep-ph/9601354.
\ref H.\ Araki, \lq On quasifree states of CAR and Bogoliubov 
automorphisms\rq, {\it Publ. RIMS Kyoto Univ.}, {\bf 6} (1970-71),
385-442.
\ref E.\ Balslev and A.\ Verbeure, \lq States on Clifford algebras\rq,
{\it Commun. Math. Phys.} {\bf 7} (1968), 55-76.
\ref E.\ Balslev, J. Manuceau and A.\ Verbeure, \lq Representations fo 
anticommutation relations and Bogoliubov transformations\rq, {\it 
Commun. Math. Phys.} {\bf 8} (1968), 315-326.
\ref S.M.\ Bilenky and B.\ Pontecorvo, \lq Lepton mixing and neutrino 
oscillations\rq, {\it Phys. Rep.}, {\bf 41} (1978), 225-261.
\ref M.\ Blasone and G.\ Vitiello, \lq Quantum field theory of fermion 
mixing\rq, {\it Ann. of Phys.}, {\bf 244} (1995), 283-311.
\ref M.\ Blasone and G.\ Vitiello, \lq Erratum: Quantum field theory of
fermion mixing\rq, {\it Ann. of Phys.}, {\bf 249} (1995), 363-364.
\ref M.\ Blasone, P.A.\ Henning and  G.\ Vitiello,
\lq Mixing Transformations in Quantum Field Theory and Neutrino 
Oscillations\rq, hep-ph/9605335.
\ref M.\ Blasone, \lq New Results in the Physics of Neutrino
Oscillations\rq, hep-ph/9810329.
\ref M.\ Blasone, P.A.\ Henning, and  G.\ Vitiello, \lq The exact formula 
for neutrino oscillations\rq, {\it Phys.Lett. B} {\bf 451} (1999) 140-145, 
hep-th/9803157.
\ref M.\ Blasone, P.A.\ Henning, and G.\ Vitiello, \lq Green's Functions 
for Neutrino Mixing\rq, hep-ph/9807370.
\ref M.\ Blasone and G.\ Vitiello, \lq Remarks on the neutrino oscillation 
formula\rq,  {\it Phys.Rev. D} {\bf 60} (1999) 111302, hep-ph/9907382.
\ref M.\ Blasone, P.\ Jizba, and G.\ Vitiello, \lq Currents and charges
for mixed fields\rq, {\it Phys.Lett. B} {\bf 517} (2001) 471-475, 
hep-th/0103087.
\ref M.\ Blasone, A.\ Capolupo, and G.\ Vitiello, \lq Comment on \lq\lq 
Remarks on flavor-neutrino propagators and oscillation formulae\rq\rq \rq, 
hep-ph/0107183.
\ref M.\ Blasone, A.\ Capolupo, and G.\ Vitiello, \lq Understanding flavor 
mixing in Quantum Field Theory\rq,  hep-th/0107125.
\ref M.\ Blasone, A Capolupo and G.\ Vitiello, \lq Quantum field theory of 
three flavour neutrino mixing and oscillations with CP violation\rq, 
hep-th/0204184.
\ref K.\ Fujii, C.\ Habe, and T.\ Yabuki, \lq Remarks on flavor-neutrino 
propagators and oscillation formulae\rq, {\it Phys.Rev. D} {\bf 64} (2001) 
013011, hep-ph/0102001.
\ref L.\ G\aa rding and A.S.\ Wightman, \lq Representations of the
commutation and anticommutation relations\rq, {\it Proc. Nat. Acad. Sci.} 
(1954) {\bf 40}, 617-626.
\ref K.C.\ Hannabuss and D.C.\ Latimer, \lq The quantum field theory of 
fermion mixing\rq, {\it J.Phys A: Math. Gen.}, {\bf 33} (2000), 1369-1373.
\ref C-R. Ji and Yu. Mishchenko \lq The general theory of quantum field 
mixing\rq, {\it Phys. Rev. D} {\bf 65} (2002), 096015-40.
\ref E.\ Kearns, T.\ Kajita and Y.\ Totsuka, \lq Detecting massive 
neutrinos\rq, {\it Scientific American}, {\bf 281} (1999), 48-55.
\ref S.F.\ King, \lq Neutrino oscillations: Status, prospects, and 
opportunities at a neutrino factory\rq, {\it J.Phys.G: Nucl.Part. Phys.},
{\bf 27} (2001), 2149-2170.
\ref R.\ Plymen and P.\ Robinson, {\it Spinors in Hilbert space},
Cambridge 
University Press, Cambridge, 1994.
\ref R.T.\ Powers \& E.\ St\o rmer, \lq Free states of the canonical 
anticommutation relations,\rq {\it Commun. Math. Phys.} {\bf 16} (1970),
1-33. 
\bye